\documentclass[twocolumn,
showpacs,preprintnumbers,amsmath,amssymb,
floatfix]{revtex4-1}

\usepackage{graphicx}
\usepackage{amsmath,amsfonts,amssymb}
\usepackage{graphicx}
\usepackage{color}
\usepackage{bbold}
\usepackage{tabularx}

\usepackage[normalem]{ulem}
\usepackage{todonotes}
\usepackage[bookmarks,
            breaklinks,
            pdfstartview=FitH, 
            pdftitle={},
            pdfauthor={},
            pdfkeywords={}
            ]{hyperref}
\usepackage{ulem}
\hypersetup{
  colorlinks   = true,  
  urlcolor     = blue,  
  linkcolor    = blue,  
  citecolor    = red    
}

\usepackage[margin=1.in]{geometry}
\usepackage{graphicx}
\usepackage{amsmath}
\usepackage{amssymb}
\usepackage{mathtools}
\usepackage{amsthm}
\usepackage{tikz-cd}
\usepackage{subfigure}
\usepackage{booktabs}

\begin{document}

\title{Efficient three-body calculations with a two-body mapped grid method }

\author{T. \surname{Secker}}

\author{J.-L. \surname{Li}}

\author{P. M. A. \surname{Mestrom}}

\author{S. J. J. M. F. \surname{Kokkelmans}}
\affiliation{Eindhoven University of Technology, P.~O.~Box 513, 5600 MB Eindhoven, The Netherlands}

\date{\today}


\begin{abstract}

We investigate the prospects of combining a standard momentum space approach for ultracold three-body scattering with efficient coordinate space schemes to solve the underlying two-body problem. 
In many of those schemes the two-body problem is numerically restricted up to a finite interparticle distance $r_\mathrm{b}$.
We analyze effects of this two-body restriction on the two- and three-body level using pairwise square-well potentials that allow for analytic two-body solutions and more realistic Lennard-Jones van der Waals potentials to model atomic interactions.
We find that the two-body $t$-operator converges exponentially in $r_\mathrm{b}$ for the square-well interaction.
Setting $r_\mathrm{b}$ to 2000 times the range of the interaction, the three-body recombination rate can be determined accurately up to a few percent when the magnitude of the scattering length is small compared to $r_\mathrm{b}$, while the position of the lowest Efimov features is accurate up to the percent level. 
In addition we find that with the introduction of a momentum cut-off, it is possible to determine the three-body parameter in good approximation even for deep van der Waals potentials.

\end{abstract}

\maketitle

\section{Introduction}
\label{sec:I}

Three-body collisions are commonly associated with atom loss and heating in ultracold atomic gases. 
However, also interesting but subtle three-body phenomena such as the Efimov effect \cite{Efimov:1970} appear in the recombination rate and modify its behavior. 
The now fifty years old prediction of the Efimov effect was experimentally discovered only in 2006 \cite{Kraemer:2006} in an ultracold gas of cesium atoms.
Key to this breakthrough observation is the
tunability of the scattering length $a$ that parametrizes the two-body interaction strength at ultralow temperatures \cite{Chin:2010}.
The Efimov regime is determined by $|a|/r_\mathrm{vdW} \gg 1$, where $r_\mathrm{vdW}$ characterizes the range of the van der Waals attraction between the atoms.
This regime is accessible close to a Feshbach resonance where $a$ goes through a pole.

In the Efimov regime the three-body system shows universal behavior that does
not depend on the details of the two-body interaction \cite{Efimov:1970, braaten:2006, Greene:2017, Naidon:2017, D_Incao:2018}. 
Remarkably an infinite sequence of loosely bound three-body states emerges
for resonant two-body interactions.
On resonance the binding energies $E_n$ of these trimers follow the universal scaling relations $E_{n+1} / E_n = \mathrm{e}^{-2 \pi / s_0}$ with $s_0 \approx 1.00624$ for identical bosons \cite{braaten:2006, Greene:2017, Naidon:2017, D_Incao:2018}.
Those scaling relations also transfer to related quantities like the scattering lengths $a_-^{(n)}$ at which the $n$-th Efimov trimer state hits the three-body threshold and causes an Efimov resonance in the three-body recombination rate.
In the universal regime the position of Efimov features is determined by a single three-body parameter, which is often determined experimentally from the position of the lowest Efimov resonance $a_-^{(0)}$ in ultracold atomic systems.
Following the pioneering work in 2006 \cite{Kraemer:2006} the $a_-^{(0)}$ has been measured over a wide range of species \cite{Kraemer:2006,pollack:2009,Gross:2009,Gross:2010,zaccanti:2009,Wild:2012,Ferlaino:2011,Berninger:2012}.
Surprisingly many of the 
three-body parameters were found to have roughly the value $a_-^{(0)} / r_\mathrm{vdW} \approx - 9$
\cite{Kraemer:2006,Gross:2009,Gross:2010,Wild:2012,Ferlaino:2011,Berninger:2012}. 
The discovery of the origin of this van der Waals universal behavior was a major theoretical success in recent years \cite{Wang:2012,Naidon:2014a,Naidon:2014l}.
However, there are still some experimental results \cite{Roy:2013,Chapurin:2019} which pose an exception to universality. 
To describe those results complex numerical models taking also the atomic spin structure into account
are necessary \cite{Chapurin:2019}.

Advanced numerical models are also needed outside the universal Efimov regime, when $|a|/r_\mathrm{vdW} \lesssim 1$. 
In this regime three-body recombination persists to constitute a major loss mechanism in an ultracold Bose gas,
but the universal expressions fixed by a three-body parameter are no longer valid.
In addition to the total recombination rate experiments can now also reveal partial recombination rates by identifying the recombination products and are thus ranging in the realm of ultracold chemistry \cite{Haerter:2013, Wolf:2017}.
Even in the regime where $a / r_\mathrm{vdW} \approx 0 $ elastic three-body effects have recently been proposed to determine the phase diagram of a Bose-Einstein condensate \cite{Zwerger:2019,Mestrom:2019,Mestrom:2020}.

The above mentioned examples substantiate that the fast experimental progress creates a demand to advance state-of-the-art theoretical models to calculate three-body effects ranging from $a/r_\mathrm{vdW} \approx 0 $ to $|a|/r_\mathrm{vdW} \gg 1$.
Therefore new numerical approaches are needed, which allow to calculate the three-body problem in an efficient way. 
The Alt-Grassberger-Sandhas (AGS) equations \citep{Alt:1967} are one way to formulate the three-body problem. 
To solve the AGS equations numerically in momentum representation, many partial wave components of the two-body $t$-operator need to be calculated for a large number of energy points, which puts constraints on calculation time and accuracy.
Fortunately, there are many coordinate space methods  \cite{Feit:1982,Monovasilis:2007,Tal:1984,Light:1985,Fattal:1996,Willner:2004,Karman:2014} to solve the two-body problem efficiently, whose capacity has been demonstrated in
two-body scattering calculations. 
A for our purpose advantagous category of methods is based on
a Discrete Variable Representation (DVR) \cite{Light:1985} or a mapped DVR \cite{Fattal:1996,Willner:2004}. 
These approaches lead to an approximate finite dimensional matrix representation of the Hamiltonian, which can be directly used to calculate the $t$-operator via 
\begin{equation}
t(z^{2b})=V+\sum_{i} V \frac{|\psi_{i}\rangle\langle \psi_{i}|}{z^{2b}-E^{2b}_{i}} V, \label{tintro}
\end{equation}
where $|\psi_{i}\rangle$ and $E^{2b}_{i}$ are eigenvectors and eigenvalues of the Hamiltonian matrix, respectively and $V$ denotes the pairwise interaction potential.
It should be noted that the diagonalization of the Hamiltonian matrix only needs to be done once, after which the $t$-operator can be calculated at any two-body energy $z^{2b}$ using Eq.~(\ref{tintro}).  

In numerical 
practice we consider a
finite relative distance between the particles 
for this two-body problem. 
The (mapped) DVR brings the free-space two-body system into a finite distance region with specific boundary conditions. 
A hard wall boundary condition is frequently chosen for some numerical benefits, especially in the mapped case \cite{Willner:2004}.
Even though the finite distance region with hard wall boundary condition has only minor impact on traditional calculations of bound state energies and wave functions, its influence on the $t$-operator which includes off-shell scattering
properties remains to be determined and is subject of study in this paper.
A study of the effect of restricting to a finite distance region on the two-body off-shell $t$-matrix and its consequences on three-body quantities is critical to clarify whether most aformentioned numerical methods \cite{Feit:1982,Tal:1984,Light:1985,Fattal:1996,Willner:2004} can facilitate the three-body calculation.

This paper is organized as follows. 
In section \ref{sec:IIA} we review the AGS equations related to three-body recombination of identical bosons interacting via pairwise interaction potentials.
In section \ref{sec:IIB} we introduce an analytic model including the finite distance region approximation with a hard wall boundary condition for the two-body $t$-operator based on a square-well interaction to analyze the validity and convergence properties of the approximation.
In section \ref{sec:IIC} we then review the mapped DVR method that we apply to a Lennard-Jones van der Waals potential with realistic long range interaction properties.
In section \ref{sec:III} we present our results for the three-body recombination rate for the square-well and Lennard-Jones potential.
We compare the results for different sizes of the finite distance region over a wide range of scattering lengths and analyze the influence of the finite distance approximation on the position of Efimov features.
Finally we analyze the convergence properties of our approach with respect to a momentum cut-off in the AGS equations for Lennard-Jones potentials that support almost 4 and 6 $s$-wave bound states.
 
\section{Theory}
\label{sec:II}

\subsection{Three-body recombination}
\label{sec:IIA}

We consider a system of three identical bosonic alkali metal atoms. 
The interaction in the system is described by pairwise interaction potentials $V_\alpha$, where the index $\alpha=(ij)$ indicates that the interaction takes place between particles $i$ and $j$.
We calculate the three-body recombination rate $K_3$ at zero kinetic energy. 
This quantity is relevant since it is directly related to the loss rate in a sample of ultracold atoms \cite{
Braaten:2008}.

To calculate $K_3$ we start from the AGS equation for three identical particles that define the transition operator $U_{\alpha 0}(z)$ related to three-body recombination into a dimer state of particles $\alpha = (ij)$ and a free particle $k$ \cite{Alt:1967,Mestrom:2019b}
\begin{align} \label{eq:AGSeqrecom}
U_{\alpha 0} (z) & = G_0^{-1}(z) \left[1 + P_+ + P_-\right]/3 \nonumber \\
& \phantom{=} + \left[ P_+ + P_- \right] \mathcal{T}_\alpha (z) G_0(z) U_{\alpha 0} (z) \, .
\end{align}
The operators $U_{\alpha 0}$, $G_0$ and $\mathcal{T}_\alpha$ depend on the complex energy $z$. $G_0$ denotes the Green's operator of the free three-body system and is defined as
\begin{equation}
G_0 (z) = (z - H_0)^{-1} \, ,
\end{equation}
whereas $\mathcal{T}_\alpha $ is related to the two-body  $t$-operator and given by
\begin{equation}
\mathcal{T}_\alpha (z) = (1 - V_\alpha G_0 (z))^{-1} V_\alpha \, .
\end{equation}
In the following we will omit the explicit dependence on $z$ for notational compactness unless it is needed.
The operators $P_+$ and $P_-$ are the cyclic and anticylclic permutation operators, respectively.

We restrict ourselves to the case of zero total angular momentum in the system, which is suitable for the low collision energy limit in an ultracold system.
For recombination from a free incoming state $\Psi_\mathrm{in}$ of energy $E$ into a $\alpha$-dimer state labeled by $d$ with wave function $\varphi_d$ and energy $E_d$ plus a free atom of absolute momentum $q_d$ relative to the dimer center-of-mass, one needs to evaluate the transition operator element
\begin{align}
& U_{\mathrm{rc}}\left( \{ \Psi_\mathrm{in} \} \to \{  q_d, \varphi_d \} \right) \nonumber \\
& \equiv {}_\alpha \langle q_d, \varphi_d | U_{\alpha 0} (z) | \Psi_\mathrm{in} \rangle
\end{align}
on the energy shell. This leads to $E = 3 q_d^2 / 4 m + E_d $ with $m$ the mass of an atom and $z = E + i0$, which means that we take the limit in $z$ from the upper half of the complex energy plane. 
In this case the relative angular momentum  between the atom and dimer in the final state is determined by the angular momentum of the dimer and the requirement that the total angular momentum needs to be zero.

We consider the limit of zero kinetic energy in the incoming state as is common for systems of ultracold atoms.
The recombination rate at zero energy is then given by \cite{Moerdijk:1995,Lee:2007,Smirne:2007}
\begin{align}
K_3 (0)
&= \frac{24 \pi m}{ \hbar} (2 \pi \hbar)^6 \nonumber \\
& \phantom{=} \sum_{d} q_d |U_{\mathrm{rc}}\left( \{ \Psi_\mathrm{in} \} \to \{  q_d, \varphi_d \} \right) |^2 \, .
\end{align}
Note that we follow the conventions of \cite{Braaten:2008} in defining $K_3$ which deviates from \cite{Moerdijk:1995,Lee:2007,Smirne:2007} by a factor of $2$.
 
The on-shell transition operator elements can be rewritten as
\begin{align}\label{eq:transopelms}
& {}_\alpha \langle q_d, \varphi_d | U_{\alpha 0} | \Psi_\mathrm{in} \rangle \nonumber \\
& = {}_\alpha \langle q_d, \varphi_d | V_\alpha G_0 \left( P_+ + P_- \right) \mathcal{T}_\alpha  G_0  U_{\alpha 0} | \Psi_\mathrm{in} \rangle \, ,
\end{align}
since the inhomogeneous term in Eq.~(\ref{eq:AGSeqrecom}) evaluates to zero in the on-shell limit and ${}_\alpha \langle q_d, \varphi_d | = {}_\alpha \langle q_d, \varphi_d | V_\alpha G_0$.
In our numerical treatment it is advantageous to consider the operator
\begin{equation}
A_\alpha = 3 G_0 \left( P_+ + P_- \right) \mathcal{T}_\alpha  G_0  U_{\alpha 0} \, .
\end{equation}
With Eq.~(\ref{eq:AGSeqrecom}) we obtain
\begin{equation}
A_\alpha = G_0 \left( P_+ + P_- \right)  \mathcal{T}_\alpha \left[ 1 + P_+ + P_- + A_\alpha \right]\, .
\end{equation}

We expand $\mathcal{T}_\alpha = \int dq q^{2}\sum_i \tau (q,i) | q, i \rangle_{\alpha \alpha} \langle q, i |$ and use the incoming state $\Psi_\mathrm{in}$, such that we arrive at the linear system 
\begin{align}\label{eq:3bodyint}
& {}_\alpha \langle q', i | A_\alpha | \Psi_\mathrm{in} \rangle  \\
& = \int dq q^{2}\sum_j {}_\alpha\langle q',i | G_0 (P_+ + P_-) | q , j \rangle_\alpha \tau (q,j) \nonumber \\
& \phantom{=} \left[ {}_\alpha \langle q, j| (1 + P_+ + P_-) | \Psi_\mathrm{in} \rangle +  {}_\alpha \langle q, j | A_\alpha | \Psi_\mathrm{in} \rangle \right] \, . \nonumber
\end{align}
The on-shell transition operator element in Eq.~(\ref{eq:transopelms}) is then directly related to the single components  ${}_\alpha \langle q_d , i | A_\alpha | \Psi_\mathrm{in} \rangle$, since the expansion base ${}_\alpha \langle q, i  |$ naturally includes terms ${}_\alpha \langle q_d, \varphi_d |V_\alpha$.
More details on the linear system can be found in appendix \ref{B}.

\subsection{Off-Shell Scattering}
\label{sec:IIB}
\begin{figure*}[htb!] 
	  \centering
	\begin{minipage}[c]{1 \columnwidth}
	  \includegraphics[width = 1 \textwidth]{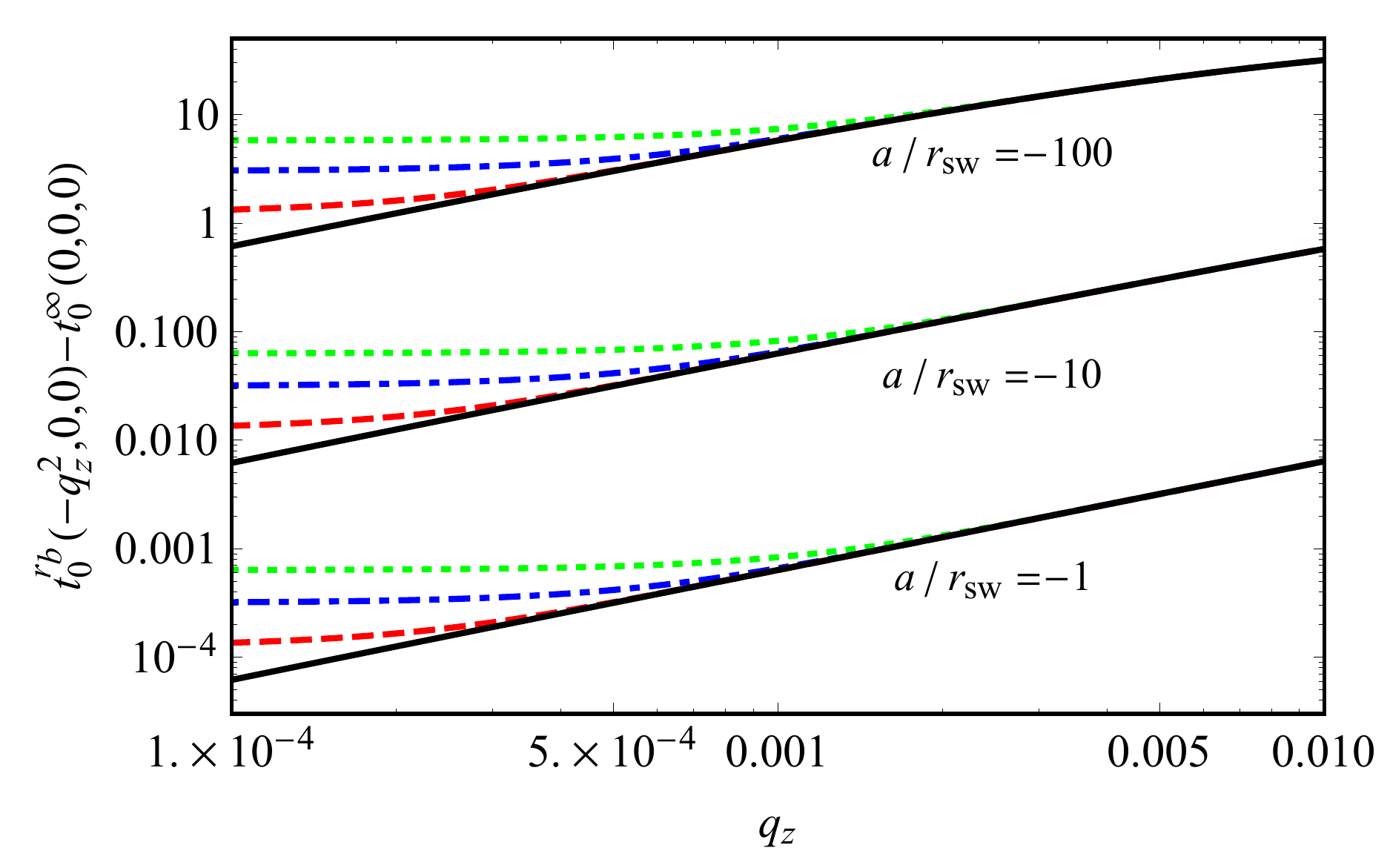}
	\end{minipage}
	\centering
	\begin{minipage}[c]{1 \columnwidth}
	\includegraphics[width = 1 \textwidth]{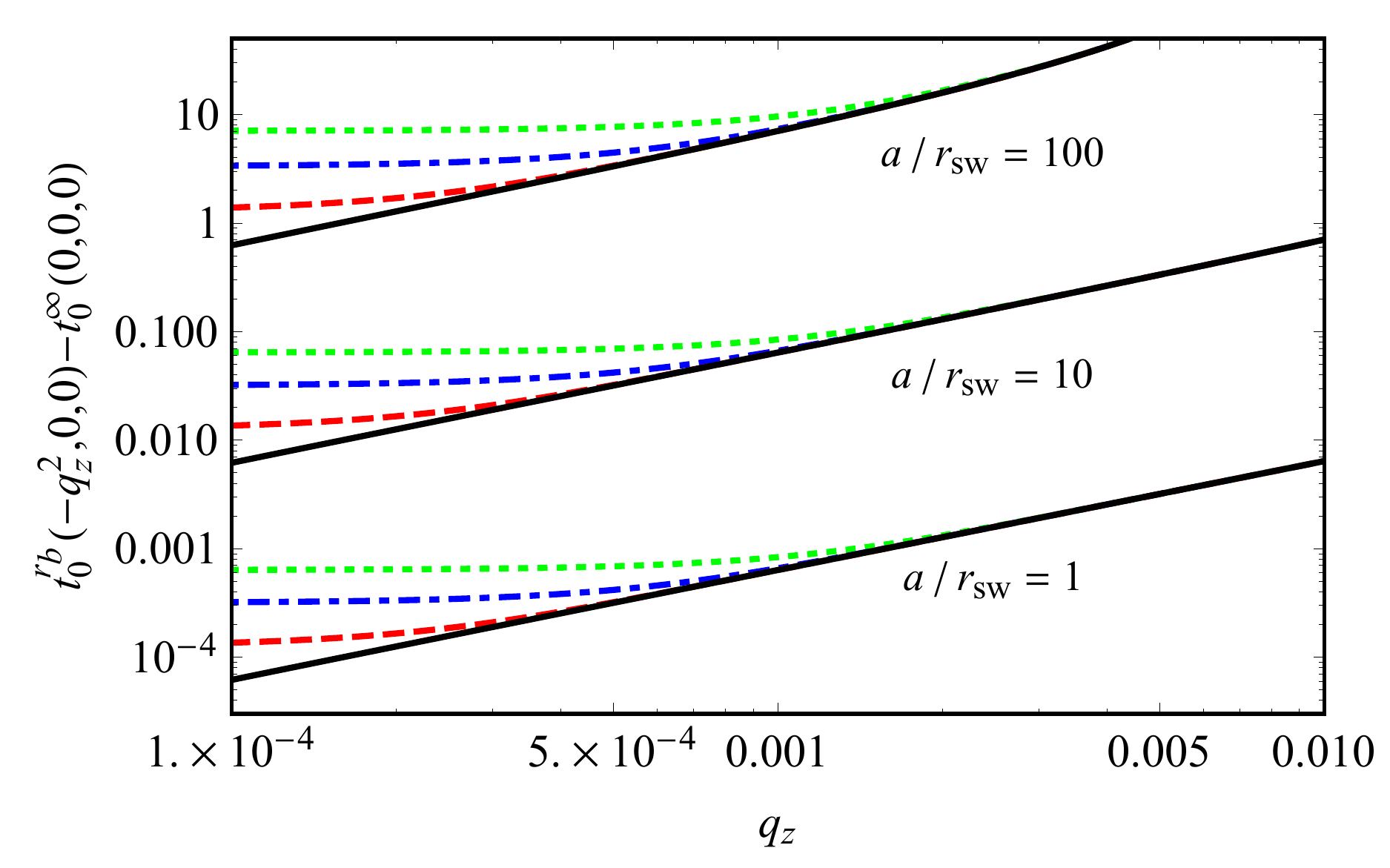}
	\end{minipage}
\caption{We compare $t_\ell^{r_\mathrm{b}}(z,0,0)$ for
$r_\mathrm{b} / r_\mathrm{sw}$ equal to $\infty$ (solid line), $5000$ (dashed line), $2000$ (dash-dotted line) and $1000$ (dotted line) for negative (left) and positive (right) scattering lengths.}
\label{fig:1}
\end{figure*}

The operator $\mathcal{T}_{\alpha}$ in the AGS equation contains all information about the interaction between the atoms. 
It is directly related to the two-body transition operator $t$ by
\begin{align}
\mathcal{T}_\alpha(z) &= \int d \mathbf{q} d \mathbf{k} d \mathbf{k}' \langle \mathbf{k}' |t(z - 3 q^2 / 4 m ) | \mathbf{k} \rangle  \\
& \phantom{=} \qquad \times |  \mathbf{k}', \mathbf{q}  \rangle_\alpha {}_\alpha  \langle  \mathbf{k}, \mathbf{q} | \, , \nonumber
\end{align}
with $\mathbf{k}$ or $\mathbf{k}'$ the relative momentum between atoms $i$ and $j$ and $\mathbf{q}$ the momentum of atom $k$ relative to the center-of-mass of the pair $\alpha=(ij)$.
We intend to compute the $t$-matrix directly using a mapped DVR 
approach described in Section \ref{sec:IIC} in combination with Eq.~(\ref{tintro}).
However, in DVR practice a finite region $[0,r_\mathrm{b}]$ with a hard wall boundary condition is introduced in the relative separation of the two atoms.
Therefore we shall analyze how well the full  $t$-operator of a two particle system can be approximated by the $t$-operator of a system with a hard wall boundary condition at a finite particle separation of $r_\mathrm{b}$.

This can be done in a clear and easy way
by analyzing a square-well interaction potential, since $t(z^{2b})$ can be worked out analytically and main features like locality and the finite range of the atomic interaction are maintained.
We define the square-well interaction potential by
\begin{equation}
V_\mathrm{sw} =
\begin{cases}
V_0, & r < r_\mathrm{sw}\\
0 , & r \ge r_\mathrm{sw} 
\end{cases} \, ,
\end{equation}
with $-V_0$ the depth and $r_\mathrm{sw}$ the range of the potential.
In case of a spherically symmetric potential $t(z^{2b})$ can be split into its partial wave components as
\begin{align}
 t(z^{2b}) 
& = \int d\mathbf{k}' d\mathbf{k} \,| \mathbf{k}' \rangle \langle \mathbf{k} |\\
& \phantom{=} \times \left[\sum_{\ell, m_\ell} \mathcal{Y}_{\ell, m_\ell}(\hat{\mathbf{k}}') t_\ell (z^{2b},k',k) \mathcal{Y}^*_{\ell, m_\ell}(\hat{\mathbf{k}}) \right]  \nonumber \, ,
\end{align}
with $\mathcal{Y}_{\ell, m_\ell}(\hat{\mathbf{k}})$ a spherical harmonic function in direction $\hat{\mathbf{k}}$.
To indicate the size of the finite distance region $r_\mathrm{b}$ we switch to the notation $ t_\ell^{r_\mathrm{b}} (z^{2b},k',k)$ with $ t_\ell^{\infty} (z^{2b},k',k)$ corresponding to the free-space case.
Changing to units where $\hbar$, the particle mass $m$ and the square-well radius $r_\mathrm{sw}$ are equal to one we get
\begin{widetext}
\begin{align} \label{eq:topsymmetric}
 t_\ell^{r_\mathrm{b}} (z^{2b}, k', k) &= \frac{2}{\pi} \frac{V_0}{k' k} \left[ \frac{-(k'^2 + k^2)(q_z^2 + q_0^2) + 2 k'^2 k^2 - 2 q_z^2 q_0^2}{2 ( k^2 - k'^2 )( k'^2 - q_0^2) (k^2 - q_0^2)} \mathrm{W}[\mathrm{S}_\ell (k r),\mathrm{S}_\ell (k' r)] \right. \\
 & \phantom{=} \left. - \frac{V_0}{2 (k'^2 - q_0^2)(k^2 - q_0^2) } \left( \frac{\mathrm{W}[\mathrm{S}_\ell(k r),  \phi_\ell^{r_\mathrm{b}}( q_z, r)] \mathrm{W}[\mathrm{S}_\ell(q_0 r), \mathrm{S}_\ell(k' r) ]}{\mathrm{W}[\mathrm{S}_\ell(q_0 r), \phi_\ell^{r_\mathrm{b}}( q_z, r))]} + k \leftrightarrow k' \right) \right]_{r = r_\mathrm{sw} = 1} \, , \nonumber
\end{align}
\end{widetext}
where the symbol $k \leftrightarrow k'$ represents the expression it is in brackets with, but with $k$ and $k'$ interchanged. We also used $q_z \equiv \sqrt{z^{2b}}$ and $q_0 \equiv \sqrt{z^{2b} - V_0} $, while $\mathrm{W}[\cdot, \cdot]$ denotes the Wronskian and $\mathrm{S}_\ell (\xi) \equiv \xi \mathrm{j}_\ell (\xi)  $ is a Riccati-Bessel function that we define via the spherical Bessel function of the first kind $\mathrm{j}_\ell$.
$\phi_\ell^{r_\mathrm{b}}( q_z, r)$ is the outer solution to the Hamiltonian differential equation $(q_z^2 - H_\ell)\phi_\ell^{r_\mathrm{b}} = 0$ with boundary condition $\phi_\ell^{\infty}( q_z, r)$ exponentially decaying as $r \rightarrow \infty$ or $\phi_\ell^{r_\mathrm{b}}( q_z, r_\mathrm{b}) = 0$ in case of finite $r_\mathrm{b}$. 
$H_\ell$ is the partial-wave two-body Hamiltonian in the relative separation $r$ between the particles. 
A derivation can be found in appendix \ref{A}.

We analyze the quality of the finite $r_\mathrm{b}$ approximation by considering the difference
\begin{align} \label{eq:topdifference}
& t_\ell^\infty(z^{2b}, k', k) - t_\ell^{r_\mathrm{b}}(z^{2b}, k', k) \nonumber \\
&= h_\ell(z^{2b}, k') h_\ell(z^{2b}, k) f_\ell(z^{2b}, r_\mathrm{b}) \, ,
\end{align}
which separates in the momenta $k, k'$ and the size of the finite distance region $r_\mathrm{b}$ (see appendix \ref{A} for more details). 
We also define the functions 
\begin{equation}
h_\ell (z^{2b}, k) = \left. \sqrt{\frac{2}{\pi}} \frac{V_0 \mathrm{W}[\mathrm{S}_\ell (q_0 r), \mathrm{S}_\ell (k r)]}{k (k^2 - q_0^2)} \right|_{r = 1} \, ,
\end{equation}
which are well behaved in $k$ for fixed $z^{2b}$
and
\begin{align}
&f_\ell (z^{2b}, r_\mathrm{b})  \\
&= \left. \frac{ - \mathrm{W}[\phi_\ell^{r_\mathrm{b}}( q_z, r), \phi_\ell^\infty( q_z, r)]}{\mathrm{W}[\mathrm{S}_\ell( q_0 r), \phi_\ell^\infty( q_z, r)]\mathrm{W}[\mathrm{S}_\ell( q_0 r), \phi_\ell^{r_\mathrm{b}}( q_z, r)]} \right|_{r = 1} \, . \nonumber
\end{align}
For $\mathrm{Im}(z^{2b}) \geq 0$ we find that the Wronskian in the numerator
\begin{equation}
\left. \mathrm{W}[\phi_\ell^{r_\mathrm{b}}( q_z, r), \phi_\ell^\infty( q_z, r)] \right|_{r = 1} \propto \mathrm{k}_\ell(r_\mathrm{b} \sqrt{-z^{2b}})
\end{equation} 
is proportional to a modified spherical Bessel function $\mathrm{k}_\ell$ that behaves like $\sim \mathrm{e}^{-r_\mathrm{b} \sqrt{-z^{2b}}} /( r_\mathrm{b} \sqrt{-z^{2b}}) $ in the limit $r_\mathrm{b} \sqrt{-z^{2b}} \rightarrow \infty $ and thus guaranties
\begin{equation}
t_\ell^{r_\mathrm{b}} (z^{2b}) \xrightarrow[r_\mathrm{b} \rightarrow \infty]{} t_\ell^\infty (z^{2b})
\, 
\end{equation}  
for all $z^{2b} \notin \mathbb{R}^+_0$ with $\mathrm{Im}(z^{2b}) \geq 0$.
We can conclude that the convergence in $r_\mathrm{b}$ will be slowest for $z^{2b}$ close to $\mathbb{R}^+_0$. 
We focus on the regime close to $z^{2b} = 0$ since for zero energy three-body scattering $z^{2b} \le 0$. 
In Fig.~\ref{fig:1} we compare $t_0^{r_\mathrm{b}}(z^{2b},0,0)$ for different sizes of the finite distance region and scattering lengths between the first and second potential resonance.
We find similar good convergence properties for positive and negative scattering lengths, while the deviation between the finite distance approximation and the free-space case increases  almost quadratically with $|a|$.

\begin{figure}[htb] 
	  \centering{}%
\includegraphics[width = 1
	  \columnwidth]{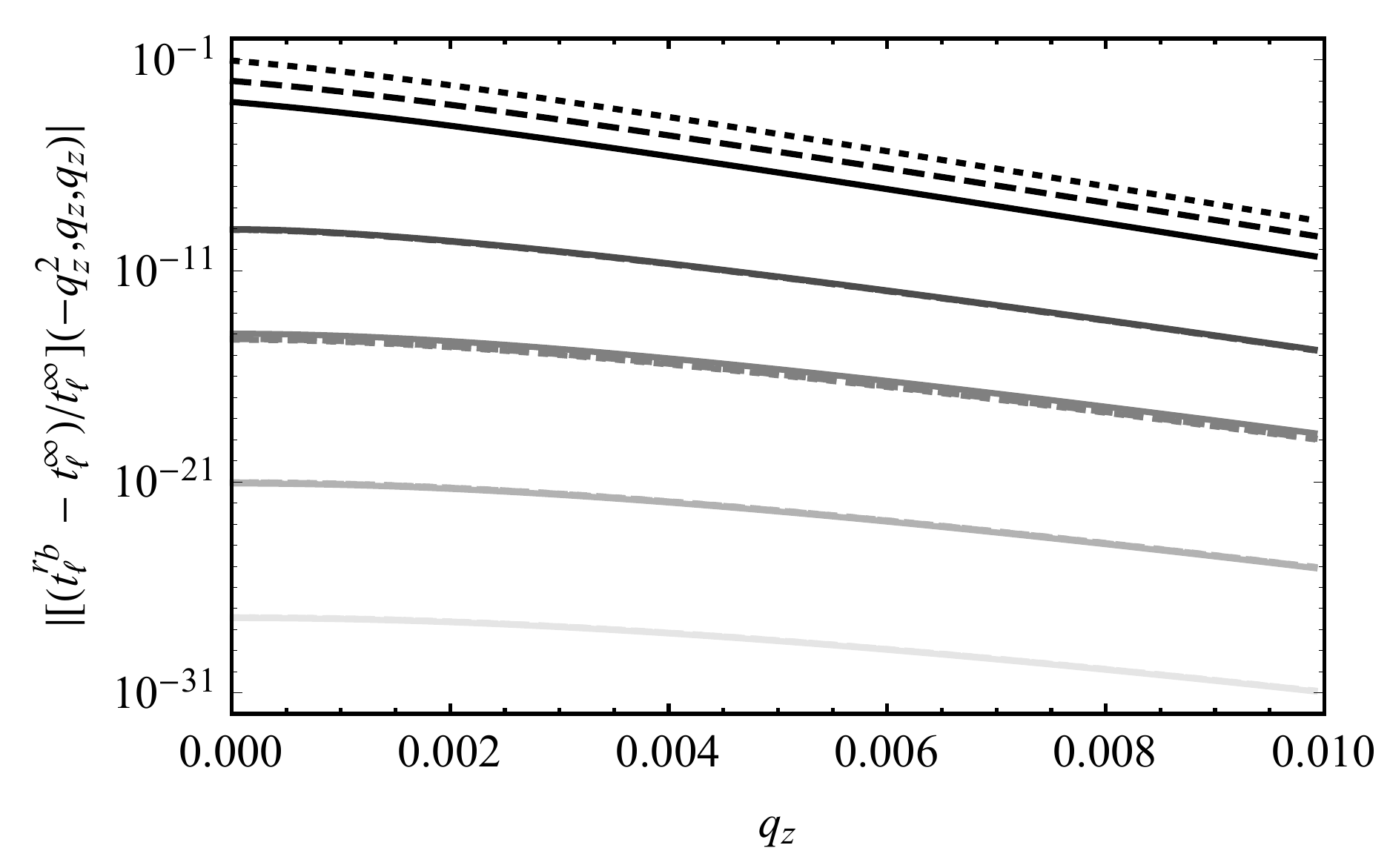}
\caption{Relative deviation of $t_\ell^\infty$ and $t_\ell^{r_\mathrm{b}}$ for $r_\mathrm{b} / r_\mathrm{sw} = 1000$ at $a / r_\mathrm{sw} = -1$ (full line), $a/ r_\mathrm{sw} = -10$ (dashed line) and $a / r_\mathrm{sw} = -100$ (dotted line) and $\ell = 0,1,2,3,4$ (black to light gray).}
\label{fig:2}
\end{figure}

In Fig.~\ref{fig:2} we show the relative deviation between the free-space and the finite distance region $t$-operator between the first and second potential resonance to quantify the quality of the approximation also for higher values of the angular momentum quantum number $\ell$.
We find that the relative deviation decreases with increasing $\ell$.
Also the expected behavior $\sim \mathrm{e}^{-r_\mathrm{b} \sqrt{-z^{2b}}} /( r_\mathrm{b} \sqrt{-z^{2b}}) $ in $q_z = \sqrt{-z^{2b}}$ can be observed.

\subsection{Mapped DVR}
\label{sec:IIC}

To compute the $t$-operator we use a mapped grid method in combination with a sine function basis as discussed in \cite{Willner:2004}.
We consider central interaction potentials with a $C_6/r^6$ van der Waals long range behavior. We focus on the Lennard-Jones van der Waals potential
\begin{equation}
V_\mathrm{LJ} = - \frac{C_6}{r^6}\left( 1 - \frac{\lambda^6}{r^6} \right) \, .
\end{equation}
The range of the van der Waals attraction is then defined as $r_\mathrm{vdW} = (m \, C_6 / \hbar^2)^{1/4} / 2 $ \cite{Chin:2010}.
For energies close to threshold the wave function will oscillate with a very big wavelength at large separation, where the potential almost vanishes and with short wavelength close to the potential minimum.
This wide range of wavelengths can increase the numerical cost when using standard grid representation methods.
Therefore we want to transform to a new relative coordinate $x$ in which the wave function oscillates with a more regular frequency. 
In the following we drop the index in the interaction potential $V_\mathrm{LJ}$, since the procedure applies generally also to potentials of similar shape.
A coordinate transformation with the desired properties is given by \cite{Willner:2004}
\begin{equation}
x(r) = \frac{\sqrt{m}}{p_\mathrm{max}} \int_{r_\mathrm{in} = 0}^r \mathrm{d}r' \sqrt{E_\mathrm{max} - V^\mathrm{env}(r')} \, ,
\end{equation}
where $p_\mathrm{max}$ and $E_\mathrm{max}$ are specifying the transformation and with the enveloping potential $V^\mathrm{env}$ whose value is at all separation lower or equal than the one of the potential $ V $ and defined as
\begin{equation}
V^\mathrm{env}(r) = \mathrm{min}_{r' \geq r}\left(V (r') \right) \, .
\end{equation}
To see that the transformation has the desired properties, we look at the semi-classical approximation of the $s$-wave Schr\"odinger equation for the phase function $\varphi$ after transforming coordinates
\begin{align}
\partial_x \varphi & = \pm \frac{\sqrt{m [E - V(r(x))]}}{\hbar (\partial_r x(r))} \\
& =  \pm \frac{p_\mathrm{max} \sqrt{ E - V(r(x))}}{ \sqrt{ E_\mathrm{max} - V^\mathrm{env}(r(x))}} \nonumber
\end{align}
which is indeed approximately constant for $E \sim E_\mathrm{max}$ and $V^\mathrm{env}(r(x)) \sim V(r(x))$.
We absorb the change in volume element $\partial_x r(x) = J(x)$ into the wave function $\langle x | \psi \rangle = \bar{\psi} (x) = \sqrt{J(x)} \psi(r(x))$ and choose an equally spaced grid $x_i = i x(r_\mathrm{b})/N  $ in $x$ with $N+1$ grid points in the region $[0,x(r_\mathrm{b})]$ together with a sine or particle in a box base 
\begin{equation}
s_k(x) = \sqrt{\frac{2}{N}}\mathrm{sin}(k \pi x / x(r_\mathrm{b}))
 \quad (k = 1, ... N-1)
 \end{equation}
on the grid.
This choice for the bases introduces a hard wall boundary condition in the system.
We define the transformation to the base of $x$-grid points 
\begin{equation}
S_{jk} = s_k (x_j) = \langle x_j | s_k \rangle \, ,
\end{equation}
which leads us to the representation of the $x$-grid points $x_j$ in terms of the sine base
\begin{equation}
\tilde{s}_j(x) = \sum_k (S^\dag)_{jk}  s_k(x) \, .
\end{equation}
The radial part of the kinetic energy operator $T = -\hbar^2 \partial_r^2 / m$ can then be obtained in the $\tilde{s}_j$-base
\begin{align}
\langle \tilde{s}_i | T |\tilde{s}_j \rangle & = - \frac{\hbar^2}{m}\frac{\pi^2}{x(r_\mathrm{b})^2}\sum_{k = 0}^N J(x_i)^{-1/2} (D^\dag)_{ik} \nonumber \\
& \phantom{=} \qquad \times J(x_k)^{-1} D_{kj} J(x_j)^{-1/2}
\end{align}
with
\begin{equation}
D_{ij} = \frac{x(r_\mathrm{b})}{\pi} (\partial_x \tilde{s}_j)(x_i) \, .
\end{equation}

As a consequence we can obtain the radial Hamiltonian $H_\ell$ of the two-atom system as a finite dimensional matrix on the $x$-grid using the $\tilde{s}_j$-base functions
\begin{equation}
\langle \tilde{s}_i | H_\ell |\tilde{s}_j \rangle = \langle \tilde{s}_i | T |\tilde{s}_j \rangle + \delta_{ij} V_\ell (x_i) \, .
\end{equation} 
Here we use the potential including the angular momentum barrier $ V_\ell (r) = V(r) + \hbar^2 \ell(\ell +1)/ m r^2$ and note that we introduce a cut-off in $V_\ell$ at large positive energy to prevent the potential to diverge at zero distance in numerical practice.
We can find the eigenvalues $E_\ell^i$ and eigenstates $| E_\ell^i \rangle$ of the resulting matrix and the two-body $t$-operator can be obtained 
in momentum representation by
\begin{align}
t_\ell(z^{2b},k',k) &= \sum_i \langle \ell,k' | x_i \rangle V(x_i) \langle x_i | \ell, k \rangle \nonumber \\
& \phantom{=} + \sum_{ijn} \langle \ell, k' | x_i \rangle \frac{\langle x_i | V | E_\ell^n \rangle \langle E_\ell^n| V | x_j \rangle}{z - E_\ell^n} \nonumber \\
& \phantom{=} \qquad \times \langle x_j | \ell, k \rangle \, , 
\end{align}
with $\langle x_i | \ell,  k \rangle = [2 J(x_i) x(r_\mathrm{b})/\pi N]^{1/2} r(x_i) \mathrm{j}_\ell [k r(x_i)] $.

Diagonalizing $t_\ell(z - 3 q_\alpha^2/4m,k',k)$ for a discrete set of momenta $k$, $k'$
we can find the expansion 
$\mathcal{T}_\alpha = \int dq q^{2}\sum_i \tau (q,i)| q, i \rangle_{\alpha \alpha} \langle q, i |$ that we need in the three-body calculation similar to \cite{Mestrom:2019b}.
In the following section we present three-body results for $V_\mathrm{LJ}$
that we obtained using the mapped DVR approach presented here.

\section{Three-body results}
\label{sec:III}
\begin{figure*}[htb!] 
	  \centering
	      \includegraphics[width=\textwidth]{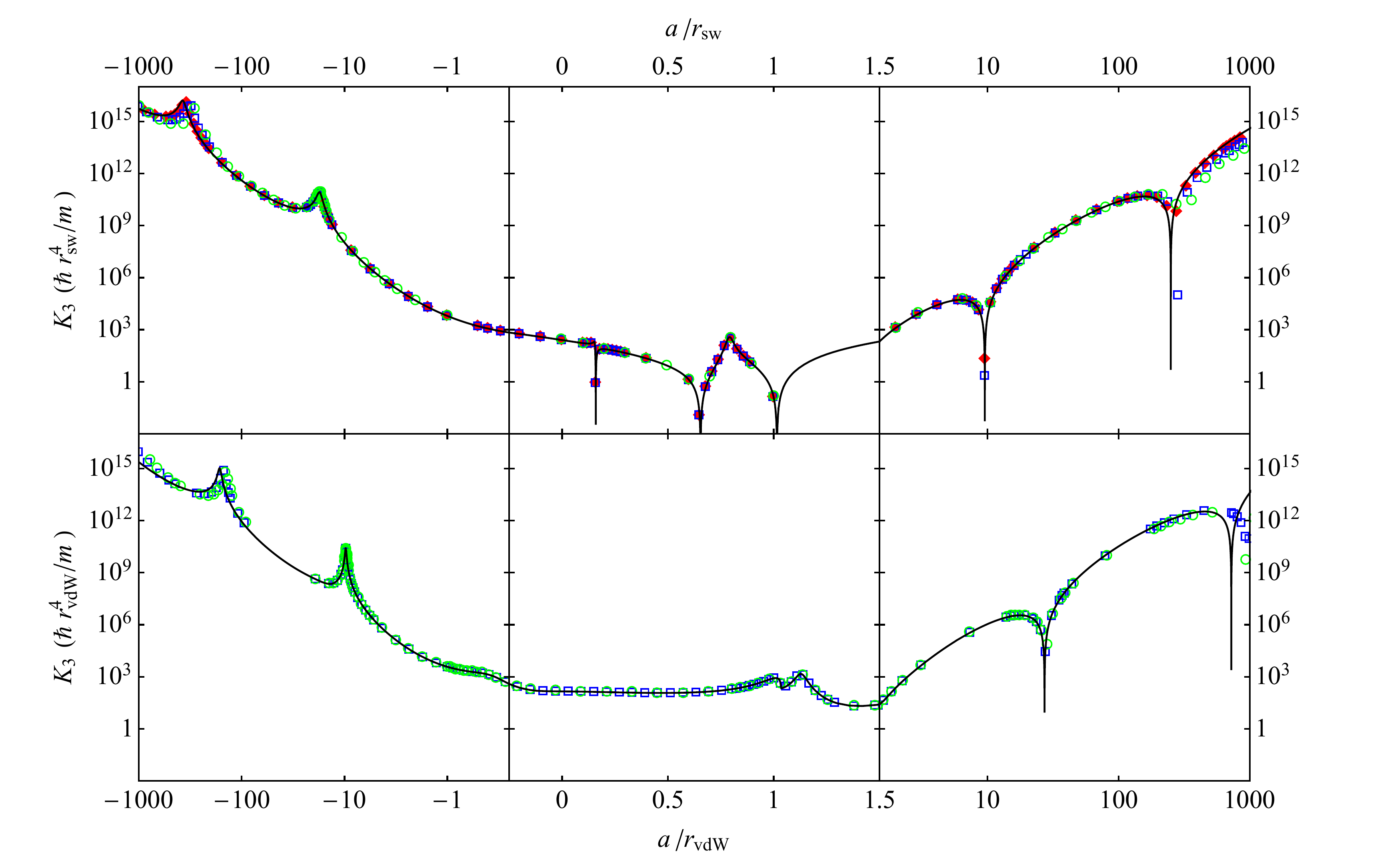}
  
\caption{\label{fig:3} Three-body recombination rates for  the square-well (top) and Lennard-Jones potential (bottom). We compare $K_3(0)$ for sizes of the finite distane region $r_\mathrm{b} / r_{\mathrm{sw}} $ or $r_\mathrm{b} / r_\mathrm{vdW} $ of $1000$ (green circles), $2000$ (blue squares), $5000$ (red diamonds),  $\infty$ (black line). The data for $r_\mathrm{b} = \infty$ are calculated using the Weinberg expansion method (see \cite{Mestrom:2019b} for more details) and have partly been published earlier \cite{Mestrom:2019,Mestrom:2020}.
}
\end{figure*}

We performed three-body recombination calculations using the analytic expressions for $t$ we obtained for the square-well interaction
and the values for $t$ that we obtained numerically for the
Lennard-Jones potential with the mapped DVR method.
In Fig.~\ref{fig:3} we compare the three-body recombination rate for different values of $r_\mathrm{b}$ over a wide range of scattering lengths $a$ between the first and second potential resonance.
The results for the free-space case $r_\mathrm{b} = \infty $ have been performed using the Weinberg expansion for the $t$-operator \cite{Mestrom:2019b} and have been partly published earlier \cite{Mestrom:2019,Mestrom:2020}.
We find that the hard wall boundary condition has a minor influence on the three-body recombination rate when $|a|/r_\mathrm{b} \ll 1$.
For the square-well interaction with $r_\mathrm{b} = 2000 \, r_\mathrm{sw}$ in the regime $|a|/r_\mathrm{sw}< 1.5$ we find a small relative deviation in $K_3$ of $\sim 1 \%$.
For the Lennard-Jones potential with $r_\mathrm{b} = 2000 \, r_\mathrm{vdW}$ in the regime $|a|/r_\mathrm{vdW} < 1.5$ we find a still small but larger relative deviation of $\sim 4 \%$ (see Fig.~\ref{fig:5} in appendix \ref{app:C}). 
We attribute this increase in relative deviation to the numerical error in the mapped DVR approach. 
However, when $|a| \sim r_\mathrm{b}$ the deviation becomes more significant.
We note that the relative deviation of the analytic $t_0$ is also increasing significantly in this regime (see Fig.~\ref{fig:2}). 

To quantify the accuracy of the finite distance approximation in the Efimov regime we investigate the ground and first excited Efimov resonance peaks and the first and second excited Efimov recombination minima for various values of $r_\mathrm{b}$.
We obtain the resonance peak positions $a_-^{(n)}$ and widths $\eta^{(n)}$ as well as the recombination minima $a_+^{(n)}$ by fitting the universal expressions \cite{Efimov:1979,Esry:1999,Bedaque:2000,Braaten:2001,Braaten:2002,Braaten:2004,braaten:2006,D_Incao:2018} for the recombination rate to our calculation close to the peak or minimum position.
We note that for fitting recombination minimum positions we introduce an overall scaling factor as an additional fit parameter to improve the quality of the fit.
The results can be found in Table \ref{tab:1} and \ref{tab:2} for different values of $r_\mathrm{b}$.
We find good agreement for both the square-well and Lennard-Jones potential in $a_-^{(0)}$ and $a_+^{(1)}$ with $r_\mathrm{b} = 2000 \, r_\mathrm{sw}$ or $r_\mathrm{b} = 2000 \, r_\mathrm{vdW}$, respectively, with a deviation of $\lesssim 1 \%$ in all cases.

\begin{table}[h!]
\centering
\caption{Width and position of the ground and first excited Efimov resonances along with first and second excited recombination minima for a square-well interaction between the first and second potential resonance with different sizes of the finite distance region.
Our results for $a_\pm^{(n)}$ are converged in at least three significant figures. 
The results for $r_\mathrm{b} / r_\mathrm{sw} = \infty $ have been obtained using the Weinberg expansion for the $t$-operator and have been partly published earlier \cite{Mestrom:2019}. \label{tab:1}}
\begin{tabularx}{\columnwidth}{>{\raggedright \arraybackslash}X|>{\centering \arraybackslash}X >{\centering \arraybackslash}X| >{\centering \arraybackslash}X >{\centering \arraybackslash}X|>{\centering \arraybackslash}X|>{\centering \arraybackslash}X}
\hline
\hline
$\frac{r_\mathrm{b}}{r_\mathrm{sw}}$ & $\eta^{(0)}$ & $\frac{a_-^{(0)}}{r_\mathrm{sw}}$ & $\eta^{(1)}$ & $\frac{a_-^{(1)}}{r_\mathrm{sw}}$ & $\frac{a_+^{(1)}}{r_\mathrm{sw}}$ & $\frac{a_+^{(2)}}{r_\mathrm{sw}}$ \\
\hline
 $1000$ & $0.059$ & $-17.12$ & $0.062$ & $-279.4$ & $-$ & $323$ \\
 $2000$ & $0.060$ & $-17.27$ & $0.049$ & $-314.6$ & $9.58$ & $284$ \\
 $4000$ & $0.061$ & $-17.34$ & $-$ & $-$ & $-$ & $-$ \\
 $5000$ & $0.061$ & $-17.36$ & $0.058$ & $-346.6$ & $9.55$ & $263$ \\
 $\infty$  & $0.061$ & $-17.42$ & $0.067$ & $-371.9$ & $9.530$ & $249.7$ \\
\hline 
\hline
\end{tabularx}
\end{table}

\begin{table}[h!]
\centering
\caption{Width and position of the ground and first excited Efimov resonances along with first and second excited recombination minima for a Lennard-Jones interaction between the first and second potential resonance with different sizes of the finite distance region. Our results for $a_\pm^{(n)}$ are converged in at least two significant figures. 
The results for $r_\mathrm{b} / r_\mathrm{vdW} = \infty $ have been obtained using the Weinberg expansion for the $t$-operator and have been partly published earlier \cite{Mestrom:2020}. \label{tab:2}}
\begin{tabularx}{\columnwidth}{>{\raggedright \arraybackslash}X|>{\centering \arraybackslash}X >{\centering \arraybackslash}X| >{\centering \arraybackslash}X >{\centering \arraybackslash}X|>{\centering \arraybackslash}X|>{\centering \arraybackslash}X}
\hline
\hline
$\frac{r_\mathrm{b} }{r_\mathrm{vdW}} $ & $\eta^{(0)}$ & $\frac{a_-^{(0)}}{r_\mathrm{vdW}}$ & $\eta^{(1)}$ & $\frac{a_-^{(1)}}{r_\mathrm{vdW}}$ & $\frac{a_+^{(1)}}{r_\mathrm{vdW}}$ & $\frac{a_+^{(2)}}{r_\mathrm{vdW}}$ \\
\hline
 $500$ & $0.022$ & $-9.46$ & $-$ & $-$& $-$ & $-$ \\
 $1000$ & $0.022$ & $-9.55$ & $0.030$ & $-139$ & $27.9$ & $954$ \\
 $2000$ & $0.021$ & $-9.59$ & $0.032$ & $-150$ & $27.5$ & $968$ \\
 $\infty$  & $0.020$ & $-9.67$ & $0.038$ & $-163$ & $27.2$ & $722$ \\
 $\infty$ \cite{Mestrom:2017}  & $-$ & $-9.74$  & $-$ & $-164$  & $27.2$  & $-$ \\
\hline 
\hline
\end{tabularx}
\end{table}

\begin{figure*}[htb] 
	  \centering
	\begin{minipage}[c]{1 \columnwidth}
	  \includegraphics[width = 1 \textwidth]{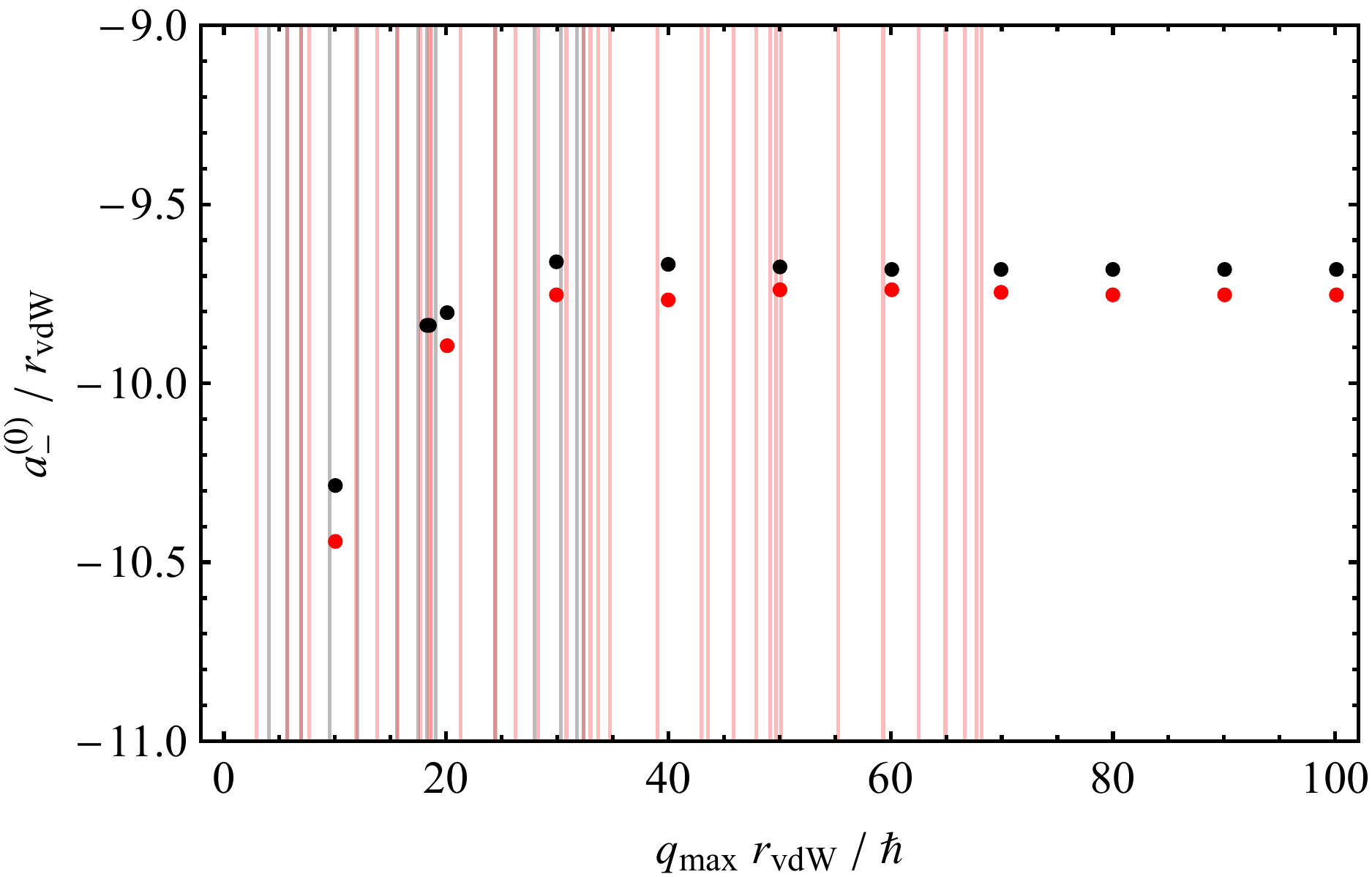}
	\end{minipage}
	\begin{minipage}[c]{1 \columnwidth}
	\includegraphics[width = 1 \textwidth]{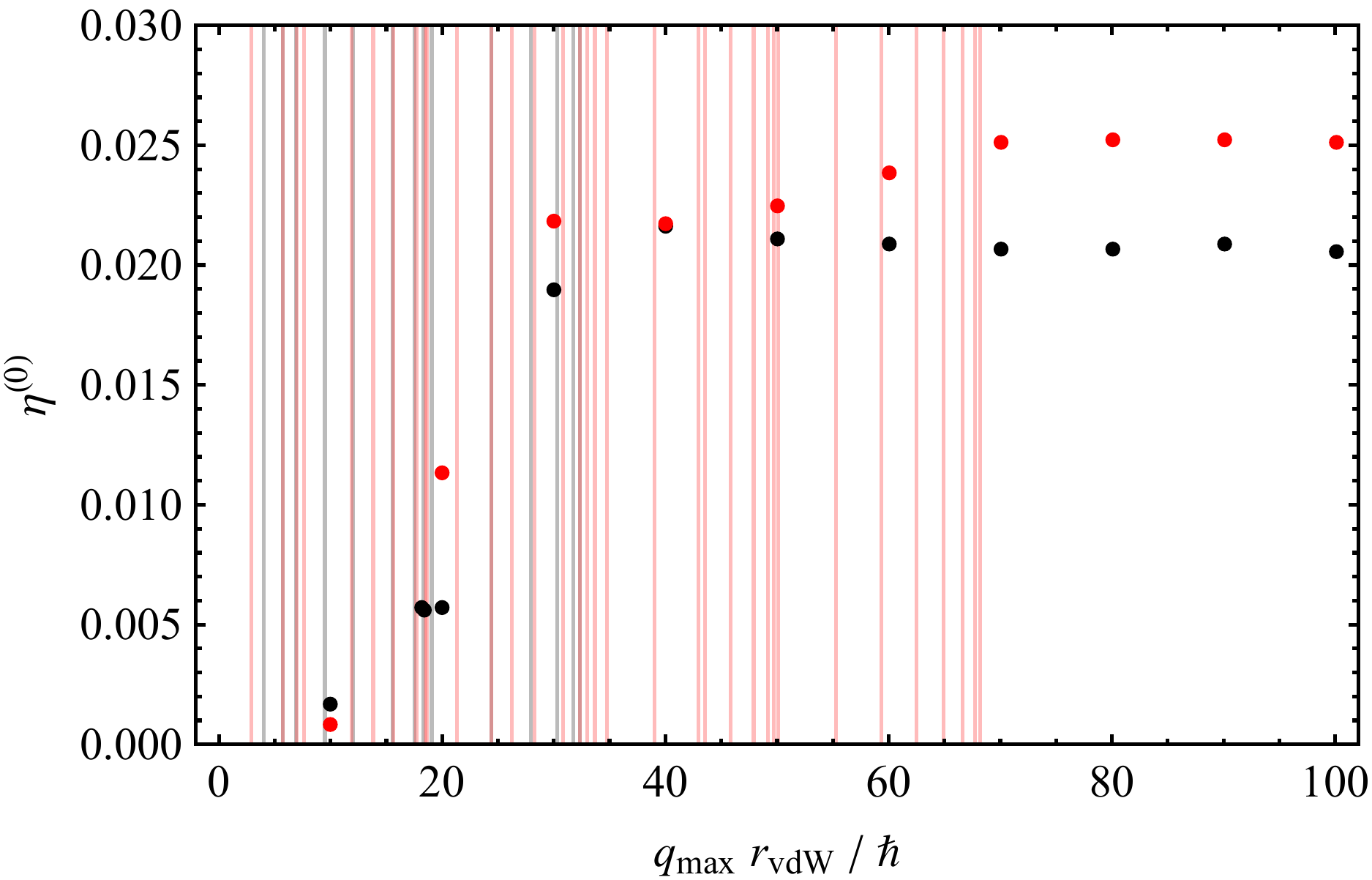}
	\end{minipage}
\caption{ We show the dependence of the lowest Efimov resonance position $a_-^{(0)}$ and width $\eta^{(0)}$ on the integration range of the atom-dimer momentum $[0,q_\mathrm{max}]$ for a Lennard-Jones van der Waals potential with almost four $s$-wave bound states (black dots) and almost 6 $s$-wave bound states (red dots). The vertical lines indicate the positions of two-body bound states for almost four (gray) and 6 (red) 
$s$-wave bound states when the system is tuned close to the lowest Efimov resonance. }
\label{fig:4}
\end{figure*}

The mapped DVR approach allows us to obtain $t$ even for deep potentials without any complications.
However, in the equations for three-body recombination, Eq.~(\ref{eq:3bodyint}), we have to limit to a finite integration range $[0, q_\mathrm{max}]$ in $q$ and to a finite expansion in relative atom-dimer partial waves $\ell$. 
In the following we consider the lowest Efimov resonance of $V_\mathrm{LJ}$ close to the 4th and 6th potential resonance. 
We include up to $\ell_\mathrm{max} = 20$ to guarantee the convergence in $\ell$
and analyze the convergence in $q_\mathrm{max}$.
The results are shown in Fig.~\ref{fig:4}.
We find that the convergence in $q_\mathrm{max}$ of $a_-^{(0)}$ is similar in both cases. 
For a value of $a_-^{(0)}$ accurate up to $2\%$ it is sufficient to choose $q_\mathrm{max} \approx  20 \, \hbar / r_\mathrm{vdW}$.
For $q_\mathrm{max} \approx  40 \, \hbar / r_\mathrm{vdW}$ we find $a_-^{(0)}$ to be accurate up to $0.2\%$. 
The resonance width $\eta^{(0)}$ is however converging more slowly in $q_\mathrm{max}$. 
To obtain converged results we find that $q_\mathrm{max}$ needs to be chosen such that $3 q_\mathrm{max}^2 / 4 m$ is larger than the lowest two-body binding energy. 
We note that when $3 q_\mathrm{max}^2 / 4 m$ is lower than some dimer binding energies recombination into those dimer states will be neglected.
The values of $a_-^{(0)}$ for the 4th and 6th potential resonance of a Lennard-Jones potential have been calculated earlier by an adiabatic hyperspherical coordinate approach in \cite{Wang:2012} to be $a_-^{(0)} /r_\mathrm{vdW} \approx -10.7$ and $a_-^{(0)}/r_\mathrm{vdW} \approx -10.4 $ respectively. 
Comparing to our results we find a relative deviation of about  $\sim 10\%$. 
We checked the convergence of our results in momentum grids, partial waves, separable expansion terms of the $t$-operator and the number of two-body $x$-grid points, but were not able to explain this deviation.
We hope to be able to resolve this discrepancy in future work.

\section{Conclusion and Outlook}
\label{sec:IV}

We extended a well established mapped DVR method to calculate off-shell scattering properties of a two-atom system that can be used directly as input for three-body scattering calculations in an AGS momentum space approach.
Using an analytic example we demonstrated that results from a method restricting to a finite distance region with a hard wall boundary condition such as the mapped DVR can be used to approximate the free space result on the two-body level.
We performed three-body recombination calculations using the mapped DVR and compared them with results calculated with a standard Weinberg expansion of the $t$-operator. 
The corresponding results are in good agreement when the magnitude of the scattering length is much smaller than the size of the finite distance region in the mapped DVR scheme.
Using the mapped DVR method we are able to perform three-body calculations even for deep interaction potentials that are more difficult to access with standard momentum space treatments.
We find that applying a cut-off in the relative atom-dimer momentum for the three-body recombination equations can lead to accurate results within a few percent when determining the three-body parameter $a_-^{(0)}$. 
However, this approximation is much less accurate for the Efimov resonance width $\eta^{(0)}$, since effects of recombination into dimer states beyond the integration range are neglected.

Our results for the atom-dimer cut-off momentum dependence suggest that the method can easily be generalized to deep realistic potentials without significant numerical complications.
Since mapped DVR methods have been applied to multichannel systems as well, the presented numerical procedure allows for a straightforward generalization to a multichannel version. 
So far multichannel methods including the full spin structure of the atomic system just exist for effective non-local potentials with a few separable terms \cite{Li:2019}.
For local van der Waals potentials on the other hand the spin structure is usually just approximated by effective models to be able to perform the calculations.
The method we present here allows for calculating three-body scattering amplitudes involving local potentials including the full spin-structure of the atomic system.
In addition we find that the regime $|a|/r_\mathrm{vdW} \lesssim 1$ is particularly well represented in the mapped DVR approach. 
This is promising since in this regime partial recombination rates have been determined experimentally \cite{Haerter:2013, Wolf:2017}.
Also the prospects for studying three-body elastic collisions around $|a| / r_\mathrm{vdW} \approx 0$ are good with the method presented here. The corresponding elastic cross sections affect
the phase diagram of a Bose-Einstein condensate, as has been demonstrated recently \cite{Zwerger:2019, Mestrom:2020}.

\section*{Acknowledgements}
We thank Denise Ahmed-Braun, Victor Colussi, Gijs Groeneveld, and Silvia Musolino for discussions.
This research is financially supported by the
Netherlands Organisation for Scientific Research (NWO)
under Grant No. 680-47-623 and by the Foundation for
Fundamental Research on Matter (FOM).

\clearpage

\appendix

\section{Details for the $K_3(0)$ determination} \label{B}
We start by giving the separable expansion of $\mathcal{T}_\alpha$ explicitly
\begin{align}
&\mathcal{T}_\alpha(z) \nonumber \\
 &= \int d \mathbf{q} d \mathbf{k} d \mathbf{k}' \langle \mathbf{k}' |t(z - 3 q^2 / 4 m ) | \mathbf{k} \rangle \\
& \phantom{=} \qquad \times |  \mathbf{k}', \mathbf{q}  \rangle_\alpha {}_\alpha  \langle  \mathbf{k}, \mathbf{q} |  \nonumber \\
&= \sum_{L, \ell, J, M_J}\int q^2 k^2 k'^2 d q d k  d k'  \nonumber \\
& \phantom{=} \qquad \times t_\ell(z - 3 q^2 / 4 m,k',k) \nonumber \\
& \phantom{=} \qquad \times |  L, \ell, J, M_J, k', q \rangle_\alpha {}_\alpha  \langle  L, \ell, J, M_J, k, q | \nonumber \\
&= \sum_{L, \ell, J, M_J, n}\int q^2 k^2 k'^2 d q d k  d k' \tau_{\ell,n}(z - 3 q^2 / 4 m) \nonumber \\
& \phantom{=} \qquad \times \chi^*_{\ell,n}(k',z - 3 q^2 / 4 m) \chi_{\ell,n}(k, z - 3 q^2 / 4 m)  \nonumber \\
& \phantom{=} \qquad \times |  L, \ell, J, M_J, k', q \rangle_\alpha {}_\alpha  \langle  L, \ell, J, M_J, k, q | \nonumber \\
&= \sum_{L, \ell, J, M_J, n}\int q^2 d q \tau_{\ell,n}(z - 3 q^2 / 4 m)  \nonumber \\
& \phantom{=} \qquad \times |L, \ell, J, M_J, n, q \rangle_\alpha {}_\alpha  \langle  L, \ell, J, M_J, n, q | \nonumber\\
&= \sum_{i}\int q^2 d q  \nonumber 
\tau(q,i)| q,i \rangle_\alpha {}_\alpha  \langle q,i | \, .
\end{align}
With $i$ we introduced a multiindex representing the tuple $(L, \ell, J, M_J, n)$,
where $L$, $\ell$, $J$ are the partial-wave quantum numbers of atom-dimer, dimer and total angular momenta, repectively, $M_J$ is the projection quantum number corresponding to $J$ and $n$ numbers the separable expansion terms $\chi_{\ell,n}$ with coefficients $\tau_{\ell,n}$ of $t_\ell$.
In the following we will indicate the quantum numbers belonging to the tuple $i$ with a subscript such that $i = (L_i, \ell_i, J_i, M_{Ji}, n_i)$. 
To make the notation more compact we will sometimes denote $\chi_{\ell_i,n_i}(k, z - 3 q^2 / 4 m)$ by $\chi(k;q,i)$ or $\tau_{\ell_i,n_i}(z - 3 q^2 / 4 m)$ by $\tau(q,i)$, such that the dependence on $z$ is implicit.

We rewrite Eq.~(\ref{eq:3bodyint}) as
\begin{align} \label{abar}
\bar{A}(q',i)&=\sum_{j}\int dq q^2 Z(q',i,q,j)\tau(q,j)\\
& \phantom{=} \times [\bar{A}_{0}(q,j)+\bar{A}(q,j)], \nonumber
\end{align}
where $\bar{A}(q,i)={}_{\alpha}\langle q, i|A_{\alpha}|\Psi_{\rm{in}}\rangle$, $Z(q',i,q,j)={}_\alpha \langle q',i | G_0 (P_+ + P_-) | q,j\rangle_{\alpha}$ and $\bar{A}_{0}(q,j)={}_{\alpha}\langle q,j| (1 + P_+ + P_-) | \Psi_{\rm{in}} \rangle$. Equation (\ref{abar}) is a Fredholm equation of the second kind,  which can be sloved by standard numerical recipes \cite{press:1996} when $Z(q',i,q,j)$, $\tau(q,j)$  and $\bar{A}_{0}(q,j)$ are known.
Since we work at zero three-body collision energy, $J$ and $M_J$ can be set to zero and  $(L \ell JM_{J})$ can be restricted to $(\ell\ell 0 0)$.
For indentical bosons $(\ell=0,2,4,\cdots)$ in general. 
For incoming states of three free atoms, $\ell=0$, such that we can write
 \begin{equation}
 |\Psi_{\rm{in}} \rangle=|L=0, \ell=0, J=0, M_J=0, k=0, q=0 \rangle,
 \end{equation}
Thus $\bar{A}_{0}(q,j)$ is expressed as
 \begin{eqnarray}
\bar{A}_{0}(q,j) = \frac{3\delta(q)}{q^2}\chi(0;q,j) \delta_{(L_j \ell_j J_j M_{Jj}),(0000)} ,
 \end{eqnarray}
and $Z(q',i,q,j)$ is given by
\begin{widetext}
\begin{eqnarray}
Z(q',i,q,j)&=&\frac{(-1)^{\ell_j}\sqrt{2\ell_i+1}\sqrt{2\ell_j+1}}{2}\int_{-1}^{1}du P_{\ell_i}\left(\frac{q'^{2}/2+q'qu}{q'\sqrt{q'^{2}/4+q^{2}+q'qu}}\right) P_{\ell_j}\left(\frac{q^{2}/2+q'qu}{q\sqrt{q^{2}/4+q'^{2}+q'qu}}\right)\nonumber \\ 
&\times& \frac{\chi^*(\sqrt{q'^{2}/4+q^{2}+q'qu};q',i)\chi(\sqrt{q^{2}/4+q'^{2}+q'qu};q,j)}{-q'^{2}/m-q^{2}/m-q'qu/m}  \label{ZZ}
\end{eqnarray}
\end{widetext}
where  $P_\ell$ is the Legendre polynomial.

\section{Derivation of the square-well $t$-operator}
\label{A}
Our derivation is closely related to that presented in Ref. \cite{Cheng:1990}.
In the following we set $r_\mathrm{sw} = m = \hbar = 1$ and denote the complex two-body energy $z^{2b}$ simply with $z$ for notational convenienc.
We want to find an expression for $t_\ell (z,k',k)$ and use the identity
\begin{equation}
t(z) = V + V g(z) V 
, 
\end{equation}
with $g(z) = (z - H)^{-1}$ the Green's operator of the relative two-body Hamiltonian $H$.
With that we arrive at
\begin{widetext}
\begin{align}
t_\ell (z,k',k) &= \frac{2}{\pi} \int d r \, r^2 \mathrm{j}_\ell (k' r) V(r) \mathrm{j}_\ell (k r) + \frac{2}{\pi} \int d r' d r \, r' \mathrm{j}_\ell (k' r') V (r') g_\ell (z, r', r) V(r) r \mathrm{j}_\ell (k r) \\
& = \frac{2}{\pi k k'} \left[ \int d r \, \mathrm{S}_\ell (k' r) V(r) \mathrm{S}_\ell (k r) + \int d r' d r \, \mathrm{S}_\ell (k' r') V (r') g_\ell (z, r', r) V(r) \mathrm{S}_\ell (k r) \right] \, , \nonumber
\end{align}
\end{widetext}
with $g_\ell (z, r', r)$ the kernel of the Green's operator $g_\ell (z) = (z - H_\ell)^{-1}$ of the radial partial wave component 
\begin{equation}
H_\ell = - \partial^2_r + \frac{\ell(\ell+1)}{r^2} + V_{\mathrm{sw}}(r)
\end{equation} 
of the Hamiltonian in the relative coordinate.
We realize that 
\begin{equation}
\int d r \, g_\ell (z, r', r) V(r) \mathrm{S}_\ell (k r) = \phi(z,k,r')
\end{equation}
is nothing but a solution to the inhomogeneous differential equation
\begin{equation}
(z - H_\ell)\phi_\ell(z,k,r) = V(r) \mathrm{S}_\ell (k r) 
\end{equation}
that vanishes for $r = 0$ and $r \rightarrow \infty$.
For the square-well potential a solution for the inner part $r \le r_\mathrm{sw}$ with correct boundary condition at $r = 0 $ is then given by
\begin{align}
&\phi_\ell^{\mathrm{inner}}(z,k,r) \nonumber \\
&= \frac{V_0}{z - k^2 - V_0 }\left[ \mathrm{S}_\ell (k r) + C \mathrm{S}_\ell (q_0 r) \right] \, ,
\end{align}
with $q_0 = \sqrt{z-V_0}$.
$C$ is a coefficient that needs to be determined by matching to the solution in the outer region $r > r_\mathrm{sw}$ where the differential equation is homogeneous since the potential $V_\mathrm{sw} = 0$ vanishes
\begin{equation}
(z - H_\ell)\phi_\ell^{r_\mathrm{b}} = 0 \, ,
\end{equation}
with boundary condition $\phi_\ell^{\infty}( q_z, r)$ decaying as $r \rightarrow \infty$ or $\phi_\ell^{r_\mathrm{b}}( q_z, r_\mathrm{b})) = 0$ in case of finite $r_\mathrm{b}$.
The matching condition at $r = r_\mathrm{sw}$ is realized when 
\begin{equation}
\left. \mathrm{W}\left[ C \mathrm{S}_\ell (q_0 r) + \mathrm{S}_\ell (k r) , \phi_\ell^{r_\mathrm{b}}(q_z,r) \right] \right|_{r = r_\mathrm{sw}} = 0 \, ,
\end{equation}
which leads to
\begin{equation}
C = \left. - \frac{\mathrm{W}\left[  \mathrm{S}_\ell (k r) , \phi_\ell^{r_\mathrm{b}}(q_z,r) \right]}{\mathrm{W}\left[ \mathrm{S}_\ell (q_0 r) , \phi_\ell^{r_\mathrm{b}}(q_z,r) \right]} \right|_{r = r_\mathrm{sw}} \, .
\end{equation}
The resulting expression for $t_\ell(z,k',k)$ can be further simplified with the integral identity
\begin{equation}
\int_0^x d r \, \mathrm{S}_\ell (a r) \mathrm{S}_\ell (b r) = \left. \frac{\mathrm{W}\left[  \mathrm{S}_\ell (b r) , \mathrm{S}_\ell (a r) \right]}{a^2 - b^2} \right|_{r = x}
\end{equation}
such that we arrive at
\begin{widetext}
\begin{align}
&t_\ell (z,k',k) \\
& = 
\frac{2}{\pi k k'} \left[ V_0 \frac{\mathrm{W}\left[  \mathrm{S}_\ell (k' r) , \mathrm{S}_\ell (k r) \right]}{k^2 - k'^2}
 + \frac{V_0^2}{z - k^2 - V_0} \left( 
\frac{\mathrm{W}\left[  \mathrm{S}_\ell (k' r) , \mathrm{S}_\ell (k r) \right]}{k^2 - k'^2} + C \frac{\mathrm{W}\left[  \mathrm{S}_\ell (k' r) , \mathrm{S}_\ell (q_0 r)  \right]}{q_0^2 - k'^2} \right) \right]_{r = r_\mathrm{sw} = 1}  \nonumber
\end{align}
\end{widetext}
To bring $t_\ell$ in symmetric form in $k$ and $k'$ and to analyze the difference in $t_\ell^{r_\mathrm{b}}$ with differing $r_\mathrm{b}$ we use the Pl\"ucker identity
\begin{align}
0 &= 
\mathrm{W}\left[ f_1 , f_2 \right]\mathrm{W}\left[ f_3 , f_4 \right] \nonumber \\
& \phantom{=} + \mathrm{W}\left[ f_1 , f_3 \right]\mathrm{W}\left[ f_4 , f_2 \right] \\
& \phantom{=} + \mathrm{W}\left[ f_1 , f_4 \right]\mathrm{W}\left[ f_2 , f_3 \right] \, . \nonumber
\end{align}
We use
\begin{align}
& \frac{1}{2}\mathrm{W}\left[ S_\ell (k r) , \phi_\ell^{r_\mathrm{b}}(q_z,r) \right] \mathrm{W}\left[ S_\ell (q_0 r) , S_\ell (k' r) \right] \\
& = - \frac{1}{2} \Big( \mathrm{W}\left[ S_\ell (k r) ,  S_\ell (q_0 r) \right] \mathrm{W}\left[ S_\ell (k' r) , \phi_\ell^{r_\mathrm{b}}(q_z,r)\right] \nonumber \\
& \phantom{=}   + \mathrm{W}\left[ S_\ell (k r) ,  S_\ell (k' r) \right] \mathrm{W}\left[\phi_\ell^{r_\mathrm{b}}(q_z,r), S_\ell (q_0 r) \right] \Big) \nonumber
\end{align}
to arrive at Eq.~(\ref{eq:topsymmetric})
and 
\begin{align}
& \frac{\mathrm{W}\left[  \mathrm{S}_\ell (k r) , \phi_\ell^{r_\mathrm{b}}(q_z,r) \right]}{\mathrm{W}\left[ \mathrm{S}_\ell (q_0 r) , \phi_\ell^{r_\mathrm{b}}(q_z,r) \right]}  
- \frac{\mathrm{W}\left[  \mathrm{S}_\ell (k r) , \phi_\ell^{r'_\mathrm{b}}(q_z,r) \right]}{\mathrm{W}\left[ \mathrm{S}_\ell (q_0 r) , \phi_\ell^{r'_\mathrm{b}}(q_z,r) \right]} \nonumber \\
& = - \frac{\mathrm{W}\left[  \mathrm{S}_\ell (k r) , \mathrm{S}_\ell (q_0 r) \right] \mathrm{W}\left[  \phi_\ell^{r'_\mathrm{b}}(q_z,r) , \phi_\ell^{r_\mathrm{b}}(q_z,r) \right]}
{\mathrm{W}\left[ \mathrm{S}_\ell (q_0 r) , \phi_\ell^{r_\mathrm{b}}(q_z,r) \right] \mathrm{W}\left[ \mathrm{S}_\ell (q_0 r) , \phi_\ell^{r'_\mathrm{b}}(q_z,r) \right]}
\end{align}
to arrive at Eq.~(\ref{eq:topdifference}).

\section{Additional figures} \label{app:C}

\begin{figure}[htb] 
	  \centering
	\begin{minipage}[c]{1 \columnwidth}
	  \includegraphics[width = 1 \textwidth]{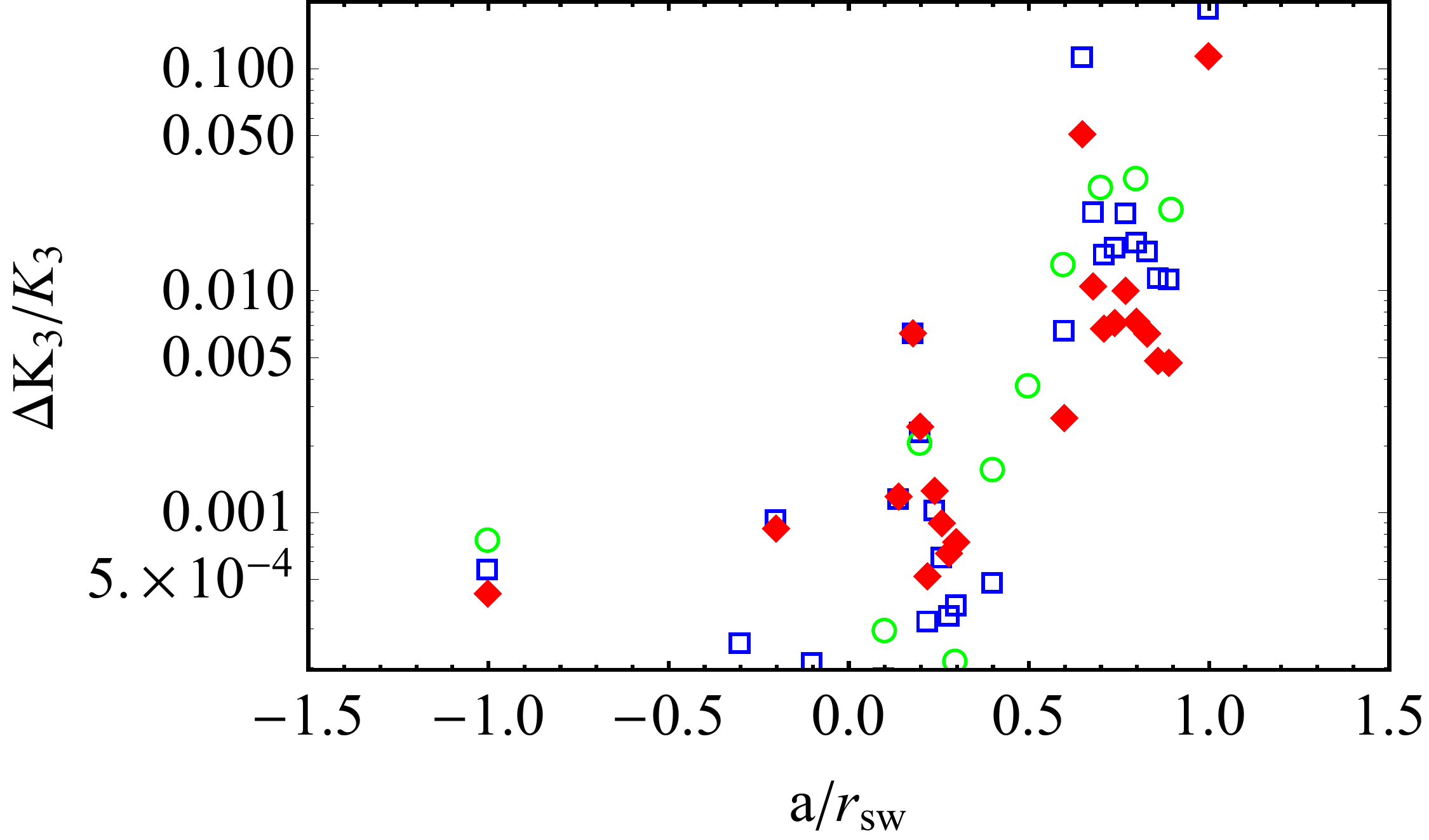}
	\end{minipage}
	\centering 
	\begin{minipage}[c]{1 \columnwidth}
	\includegraphics[width = 1 \textwidth]{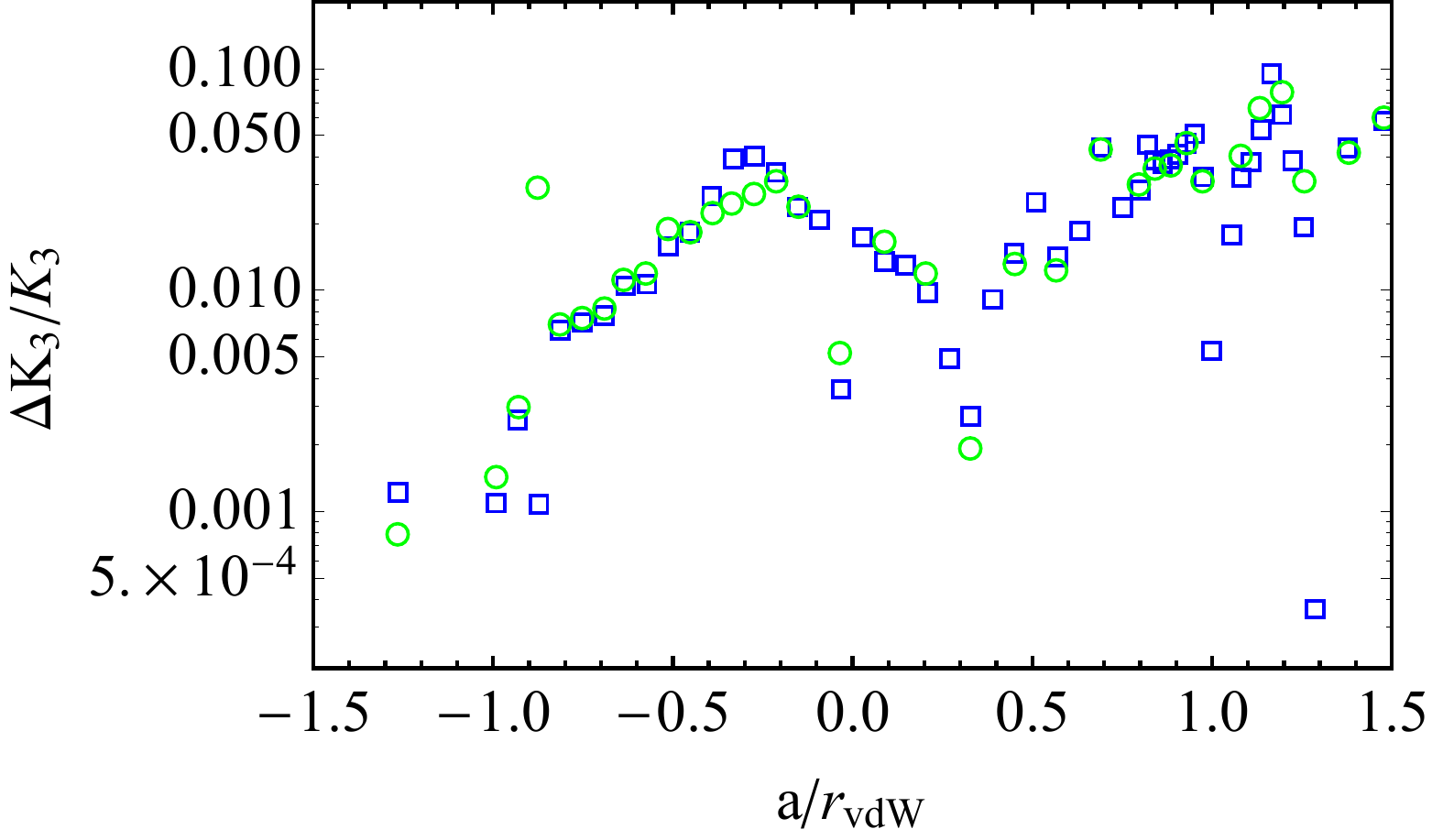}
	\end{minipage}
\caption{We show the absolute relative deviation of the finite $r_\mathrm{b}$ and free space three-body recombination rates $\Delta K_3 / K_3$ for sizes of the finite distance region $r_\mathrm{b} / r_{\mathrm{sw}} $ or $r_\mathrm{b} / r_\mathrm{vdW} $ of $1000$ (green circles), $2000$ (blue squares), $5000$ (red diamonds). The upper panel shows the square-well results and the lower one the Lennard-Jones results.}
\label{fig:5}
\end{figure}
\clearpage

\bibliography{biblio-TS}
\clearpage

\end{document}